\documentclass[twocolumn,prl,showpacs,aps,a4paper]{revtex4}

\usepackage{graphicx}
\usepackage{multirow}
\usepackage[dvipsnames]{xcolor}
\usepackage{amsmath,amsfonts,amssymb}
\usepackage{xr-hyper}
\usepackage[hidelinks]{hyperref}
\begin{document}

\title{Electron heating and mechanical properties of graphene}

\author{Jose Angel Silva-Guill\'{e}n}
\email{josilgui@gmail.com} 
\affiliation{ Key Laboratory of Artificial Micro- and Nano-structures of Ministry of Education and School of Physics and Technology, Wuhan University, Wuhan 430072, China}

\author{Francisco Guinea}
\affiliation{Fundaci\'on IMDEA Nanociencia, C/Faraday 9, Campus Cantoblanco, 28049 Madrid, Spain}
\affiliation{Department of Physics and Astronomy, University of Manchester, Oxford Road, Manchester M13 9PL, UK}
\affiliation{Donostia International Physics Center (DIPC) -- UPV/EHU, E-20018 San Sebasti\'an, Spain}

\begin{abstract}
The heating of electrons in graphene by laser irradiation, and its effects on the lattice structure, are studied. Values for the temperature of the electron system in realistic situations are obtained. For sufficiently high electron temperatures, the occupancy of the states in the $\sigma$ band of graphene is modified. The strength of the carbon-carbon bonds changes, leading to the emergence of strains, and to buckling in suspended samples. While most applications of ``strain engineering'' in two dimensional materials focus on the effects of strains on electronic properties, the effect studied here leads to alterations of the structure induced by light. This novel optomechanical coupling can induce deflections in the order of $\sim 50$ nm in micron size samples.
\end{abstract}
\maketitle

\paragraph{\textbf{Introduction}---}
Graphene, and other two dimensional materials, show a unique coupling between the electronic and mechanical properties\cite{Aetal16,RoldanTheory2017}. As a result, electronic transport and optical transitions depend on the shape of the sample, and on strains which may be present. The term ``strain engineering'' is commonly used\cite{PN09} to describe techniques which use modifications of the strains in the system to induce desirable electronic properties.  The inverse effect, the modification of structural properties by changes in the electronic structure is hampered by the high mechanical stiffness of these materials. The low optical absorption of a graphene single layer is an additional difficulty, if the changes in the electronic distribution are induced by light. Proposals for the creation of significant forces by optical means rely on non trivial combinations of graphene and dielectric layers\cite{Metal14c,SISZTM16}. Unusual effects of light on macroscopic, graphene like systems has also been reported\cite{Zetal15}.

It is well documented that intense laser pulses lead to the excitation of high energy electron-hole pairs, and, ultimately, to an electron plasma in thermal equilibrium at a temperature much higher than the lattice temperature\cite{WinzerGraphene2010,Letal10,Jetal13,Getal13,Jetal15,Metal16}. The cooling of a hot electron plasma is mediated by optical and acoustic phonons\cite{BM09,WS09}. The decoupling between electronic and lattice degrees of freedom has been studied extensively. The difference between electronic and sound velocities suppresses the phase space available for electron-phonon scattering\cite{SRL12,SL15}, and reduces the transfer of energy from electrons to phonons. This obstruction is partially relieved by coherent processes involving phonons and impurities\cite{SRL12}, named supercollisions, which have been observed in different experiments\cite{Getal11,Tetal13,STKL13,Tetal15}.

\begin{figure}
\centering
\includegraphics[scale=1]{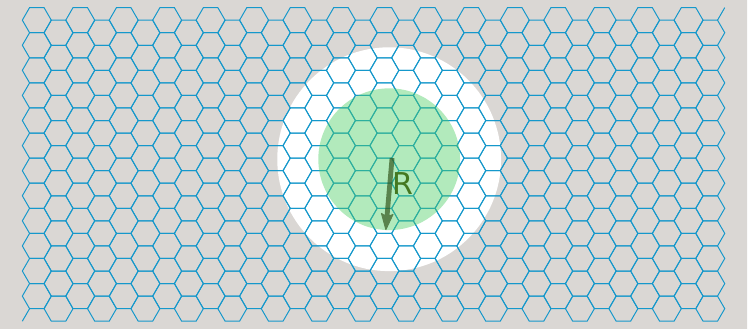}
\caption{Schematic drawing of the device used to heat up electrons in a graphene lattice. 
The substrate is depicted in gray and has a hole. A region of suspended graphene is illuminated by a laser beam (depicted in green) of radius $R$.}
\label{fig:model}
\end{figure}

In order to analyze the influence of a high temperature electron gas on the structural properties of graphene we consider the setup sketched in Fig.~[\ref{fig:model}]. A region in a suspended graphene layer of radius $R$ is illuminated by a laser. The graphene layer absorbs energy from the laser beam, at a rate defined by the laser power, $W$. We calculate the electron temperature once a steady state regime has been reached. The electron temperature modifies the occupancy of the graphene bands, which, in turn, changes the forces between atoms and induce strains and deformations in the lattice, {\it irrespective of the lattice temperature, which is assumed to be much lower than the electron temperature.} The effect of the electron temperature on the lattice constant is obtained from a self consistent band structure calculation, using a Boltzmann distribution for the occupancy of the electronic states. This calculation includes the non negligible coupling between the $\sigma$ bands and the lattice parameter\cite{Metal13,MFBHHW17}. For completion, we include a brief discussion of the radiation pressure due to the momentum transfer from laser photons to the graphene layer in the Supplementary Information\cite{si}.

\paragraph{\textbf{Electron and lattice temperature}---}

We assume that the laser energy is absorbed by the graphene layer via the creation of electron-hole pairs of energies comparable to $\hbar \nu$, where $\nu$ is the frequency of the laser. 
We study a suspended layer, and we do not need to take into account the degrees of freedom of a substrate. A steady state is reached through the combined effect of electron-electron interactions, the scattering between electrons and optical phonons, and heat diffusion, which transfers energy away from the region illuminated by the laser. The electron-electron interactions, which thermalize the electron plasma, include plasmon emission processes\cite{HPBGH16}. The steady state is characterized by a temperature $T ( W , R )$ which describes both the electron gas and the optical phonons\cite{substrate}. The temperature depends on the laser power, $W$, and the radius of the laser spot, $R$.

The rate at which the energy flows away from the illuminated area depends on the electron thermal conductivity, $\kappa_{el}$, which is given by\cite{WS09}
\begin{align}
 \kappa_{el} &   =  \frac{18 \zeta(3) k_B^3 T^2_{e-h}}{\pi^2 \hbar^2 v_F^2 a}v_F \ell_{el}, \label{eq:kappael}
 \end{align}
 where $\zeta(x)$ is the Riemann zeta function, $a\approx3.5$~\AA~is the thickness of the graphene layer, $k_B$ is the Boltzmann constant, $v_F$ the Fermi velocity, and $\ell_{el}$ is the electronic mean free path. At high temperatures, $k_B T_{e-h} \gg \mu$, where $\mu$ is the chemical potential,  the value of $\ell_{el}$ is given by
  \begin{equation}
 \ell_{el} \approx \frac{e^2}{k_B T_{e-h}} \approx \frac{\hbar v_F}{k_B T_{e-h}}.
  \label{eq:l_el}
\end{equation}
Hence, 
\begin{equation}
  \kappa_{el}\approx \frac{18 \zeta ( 3 )}{\pi^2} \frac{ k^2_B T_{e-h}}{\hbar a} = \frac{c_{\kappa} k^2_B T_{e-h}}{\hbar a},
  \label{kappaeh_f}
\end{equation}
where $c_{\kappa} \approx 18 \zeta (3)/\pi^2$ is a dimensionless constant.
This result, as well as the existence of a universal electrical conductivity, $\sigma \sim e^2/\hbar$, are a consequence of the fact that neutral graphene is a critical system.

If we take values for the thermal conductivity for the lattice from previous works, that is $\kappa_l \approx 5000\frac{\mathrm{W}}{\mathrm{m}\cdot\mathrm{K}}$\cite{Betal08,B11},
Eq. \ref{eq:kappael} implies that $\kappa_l \gg \kappa_{e-h}$, even for electron-hole temperatures $T_{e-h} \gtrsim 10^3 \textrm{K}$.
Therefore, in the following, we can assume that the lattice dissipates heat rapidly, and remains in equilibrium with the external environment.

The rate of heat flow from the electrons to the acoustic modes can be divided into two contributions, one  where the total momentum is conserved\cite{BM09}, and the other from supercollisions, which is mediated by elastic scattering\cite{SRL12}.
The heat flow rates per unit area for these processes are
\begin{align}
  \left. \frac{\partial {\cal Q}}{\partial t} \right|_{ac1} &\approx  \frac{D^2 ( k_B T_{e-h} )^5}{\hbar^5 \rho v_F^6}, \nonumber \\
    \left. \frac{\partial {\cal Q}}{\partial t} \right|_{ac1} &\approx  \frac{D^2 ( k_B T_{dis} )^2 ( k_B T_{e-h} )^3}{\hbar^5 \rho v_F^4 c^2},
 \label{rate_acoustic}
\end{align}
where $D$ is the deformation potential, $\rho$ is the mass density, and $c$ is the sound velocity. We describe supercollisions in terms of an effective temperature related to elastic scattering, $T_{dis}$. We assume that elastic scattering leads to a mobility independent of the carrier density. Then, $( k_B T_{dis} )^2 \approx \epsilon_F \cdot ( \hbar v_F / \ell_{el}^{elastic} )$, where $\ell_{el}^{elastic}$ is the electronic elastic mean free path. 
For $\ell_{el}^{elastic} \sim 100$ nm, $T_{dis} \sim 10^2$ K.

\begin{figure}
  \centering
  \includegraphics{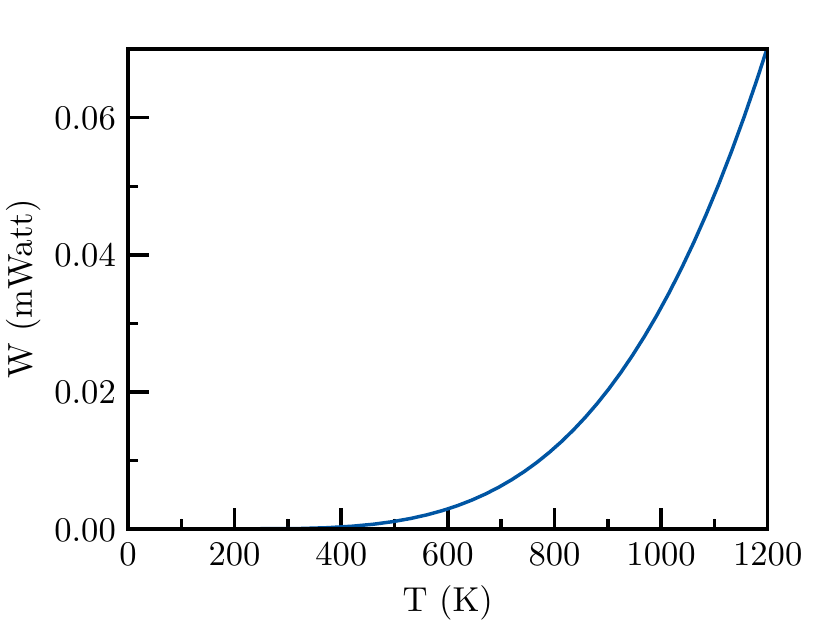}
  \caption{Power dissipated by optical phonons over an area $A = 1 \mu$m$^2$ as function of electron temperature.}
  \label{fig:optical}
\end{figure}

The rate of heat flow to the optical modes is
\begin{align}
\left. \frac{\partial {\cal Q}}{\partial t} \right|_{op} &\approx \alpha_{op} \frac{( k_B T_{e-h} )^3}{( \hbar v_F )^4} {\cal F} \left( \frac{\hbar \omega_{op}}{k_B T_{e-h}} \right),
\label{rate_optical}
\end{align}
where $\omega_{op}$ is an average optical phonon frequency, and the parameter $\alpha_{op}$ and the function ${\cal F}$ are described in the Supplementary Information (see also Ref. \cite{Letal10,YTP19}).
The power dissipated to optical modes in an area $A = 1 \mu$m$^2$ as function of electron temperature is plotted in Fig.~[\ref{fig:optical}].

By comparing  Eqs. (\ref{rate_acoustic}) and (\ref{rate_optical}), we conclude that the energy transfer from high temperature electrons to optical phonons is much larger than the energy transfer to acoustic phonons,
for physically accessible temperatures, $T_{e-h} \lesssim 10^4$ K.
Therefore, in the following, we can consider only the role of the optical phonons.

\begin{figure}
  \includegraphics{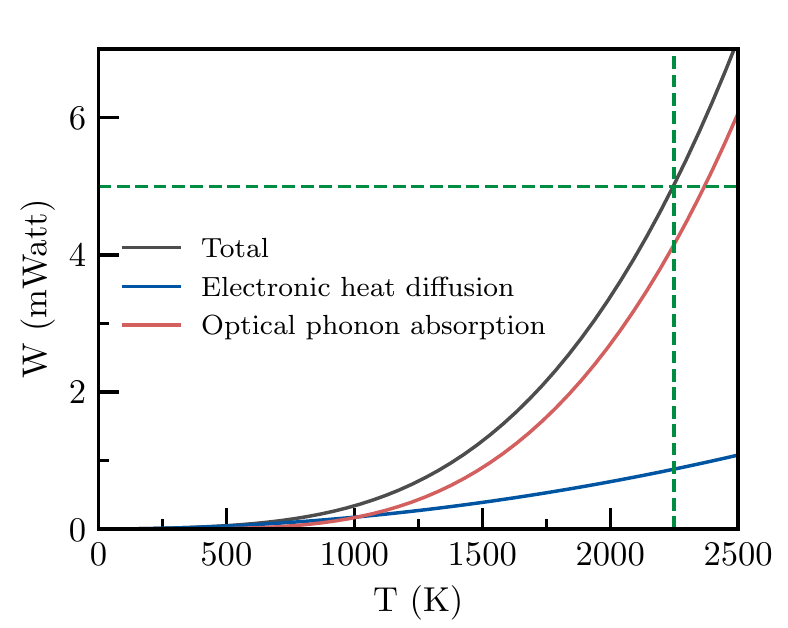}
	\caption{\label{fig:comp}
	Temperature of the electron-hole plasma versus laser power for a laser spot of radius $R = 1\mu$m. Red: Dissipation by electronic heat diffusion. Blue: Dissipation by optical phonon absorption. Black: total dissipation. For a laser power of $W = 5$ mWatt, the calculated temperature is $T = 2250$ K.}
\end{figure}

The fraction of the laser power, $W$, absorbed by the electron-hole pairs in the graphene layer is 
$\pi \alpha W$, where $\alpha \approx 1 / 137$ is the fine structure constant, and $\pi \alpha$ is the optical absorption of a graphene layer\cite{Netal08}. In order to obtain the steady state temperature of the electron plasma, we take into account the energy dissipated away from the laser spot, which depends on the electron heat conductivity, and the transfer of energy from electrons to optical phonons. We obtain
\begin{align}
c_\kappa \frac{k_B^2}{\hbar} \partial_r [ T ( r ) \partial_r T ( r ) ] &+  \frac{\alpha_{op} [ k_B T ( r ) ]^3 }{( \hbar v_F )^4} {\cal F} \left( \frac{\hbar \omega_{op}}{k_B T} \right) = \nonumber \\ &= \left\{ \begin{array}{cc} \frac{\alpha W}{\pi R^2} & r \le R \\ 0 &R < r \end{array} \right.,
\label{eq:diff}
\end{align}
where $c_\kappa$ was defined in Eq.~(\ref{kappaeh_f}). As mentioned previously, we assume that the laser has power $W$, and it irradiates uniformly a circular spot of radius $R$. Qualitatively, the
two terms in Eq.~(\ref{eq:diff})  allow us to define two cooling regimes:

- For low values of $R$, or large values of $W$, the dissipation is dominated by electronic thermal conduction into the non irradiated region, $r > R$, described by the first term in Eq.~(\ref{eq:diff}). 

- If $W$ is sufficiently low, or $R$ is large enough, dissipation is mostly the local transfer of heat to optical modes, given by the second term in Eq.~(\ref{eq:diff}). 

The values, $W_*$ and $R_*$, which define the crossover between these regimes takes place approximately,  are

\begin{align}
  \alpha W_* R_*^4 &\sim \frac{( \hbar v_F )^8}{\alpha_{op}^2 \hbar^3},
\end{align}

where we have replaced ${\cal F} [ (\hbar \omega_{op} ) / ( k_B T ) ]$ by its constant value in the limit $(\hbar \omega_{op} ) / ( k_B T ) \rightarrow \infty$ (see Supplementary Information for more details).

A more precise determination can be obtained from computing the electron temperature as function of $W$ and $R$ considering only one relaxation mechanism. 
The crossover between the two regimes takes place when the two temperatures are similar. Fig.~[\ref{fig:comp}] shows the relation between plasma temperature and laser power when the laser is focused on a region of radius $R = 1\mu$m. 
The optical phonon absorption dominates for $W \gtrsim 0.2$ mWatt. 
For this laser power, the electron plasma reaches a temperature $T \approx 1300$ K.
For a power $W \sim 5$ mWatt, in the regime dominated by optical phonons, the electron temperature is $T \approx 2250$ K.

Outside the region illuminated by the laser, electronic thermal conduction will bring the electron-hole plasma to equilibrium with the external environment. 
From Eq.~(\ref{eq:diff}) we can define a length scale
\begin{align}
\ell_{eq}^2 &= \pi R^2 \frac{c_{\kappa} ( k_B T )^2 }{\alpha W}.
\end{align}
For $T \sim 2200$ K and  $W \sim 5$ mWatt, we obtain $\ell_{eq} \sim 600$ nm.

\begin{figure}
  \centering
  \includegraphics{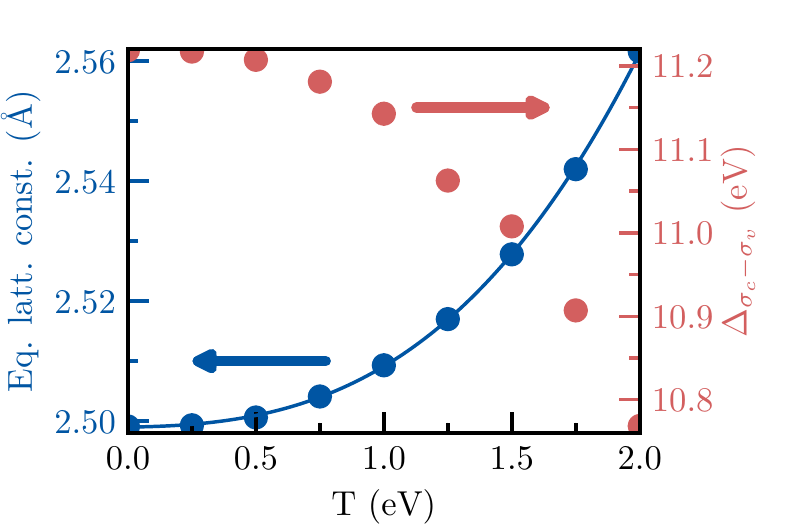}
  \caption{Equilibrium lattice constant (blue) and gap between the $\sigma$ bands (red) vs. electronic temperature.}
\label{fig:dft}
\end{figure}

\paragraph{\textbf{Effect of the electronic temperature on the graphene lattice---}}
As we have just seen, it is possible to change the temperature of electrons in graphene without modifying the actual temperature of its lattice.
Now, we center our attention on the possible consequences that the change of the electronic temperature has on the lattice of graphene.
For that, we carried out first-principles calculations.
These were performed using a numerical atomic orbitals approach to DFT,\cite{HohKoh1964,KohSha1965} which was developed for efficient calculations in large systems and implemented in the \textsc{Siesta} code.\cite{SolArt2002,ArtAng2008}
We have used the generalized gradient approximation (GGA) and, in particular, the functional of Perdew, Burke and Ernzerhof.\cite{PBE96}
Only the valence electrons are considered in the calculation, with the core being replaced by norm-conserving scalar relativistic pseudopotentials~\cite{tro91} factorized in the Kleinman-Bylander form.\cite{klby82} 
The non-linear core-valence exchange-correlation scheme~\cite{LFC82} was used for all elements. 
We have used a split-valence triple-$\zeta $ basis set including polarization functions.\cite{arsan99} 
The energy cutoff of the real space integration mesh was set to 1000 Ry. 
To build the charge density (and, from this, obtain the DFT total energy and atomic forces), the Brillouin zone (BZ) was sampled with the Monkhorst-Pack scheme\cite{MonPac76} using grids of (60$\times$60$\times$1) {\it k}-points. 
To simulate the effect of increasing the electronic temperature of graphene, we changed the electronic temperature of the Fermi-Dirac (FD) distribution of the electrons. 
It is important to note that, once a finite temperature has been chosen, the relevant energy is not the Kohn-Sham (KS) energy, but the Free energy since the atomic forces are derivatives of this.\cite{WentzcovitchEnergy1992,GironcoliLattice1995,vasp2}
The change of the lattice constant with electronic temperature is shown in Fig. Fig.~\ref{fig:dft}, where we can see that with increasing temperature, the lattice constant becomes larger.

This result can be understood looking at the effect that the electronic temperature has on the $\sigma$ bands which are the responsible for the bonds in graphene and, therefore, its lattice constant.
Looking at Fig.~\ref{fig:dft}, we can see that changing the electronic temperature slightly changes the population of these bands. 
As a result, the $\sigma$ bonds will be become weaker, and the lattice will expand.

\begin{figure}
  \includegraphics{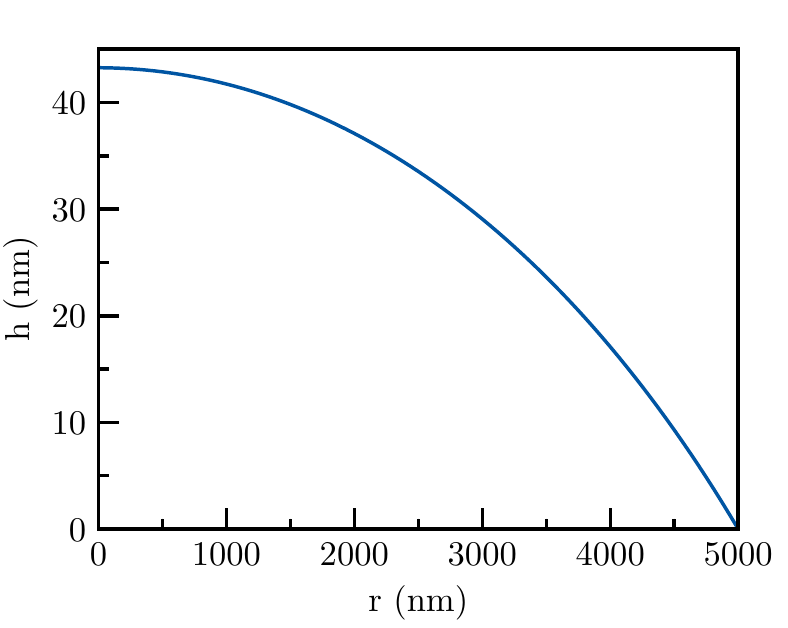}
	\caption{Profile of a disk shaped suspended graphene flake with clamped edges of radius $ R = 5000$ nm with a thermally generated stress of $\epsilon = 0.01 \%$.}\label{fig:expansion}
\end{figure}
\paragraph{\textbf{Electronic temperature and strains}---}
We have just seen that the temperature of the electron-hole plasma can modify the interatomic forces and the local lattice constant. 
Hence, strains are induced in the graphene layer when shining a laser beam to a graphene layer.

The results in the preceding paragraph suggest that a lattice expansion of order $\epsilon = \Delta \ell / \ell \sim 0.01 \%$ is possible when the temperature of the electron plasma is $T \sim 1000 - 2000$ K.
In a suspended system with clamped edges (see Fig. \ref{fig:model}), such an expansion will make the sheet to buckle. 
A simple calculation, using the techniques developed for graphene bubbles in Ref.~\cite{KGLGG16}, and for a circular region of radius $R = 5 \mu$m gives the  profile  shown in Fig.[\ref{fig:expansion}]. Note that the average strain is $\epsilon \approx h_{max}^2 / R^2$.

\begin{figure}
  \includegraphics{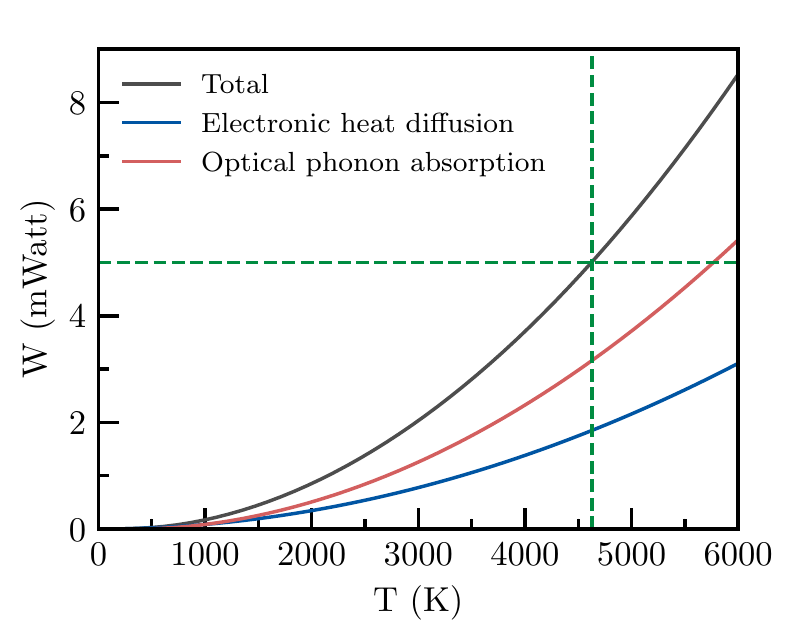}
	\caption{As in Fig.[\ref{fig:comp}] for a graphene bilayer. The calculated
temperature is $T \approx 4570$ K.}\label{fig:bilayer}
\end{figure}

\paragraph{\textbf{Generalization to multilayer graphene.}---}
{\it A graphene bilayer.}
The rate of heat transfer in a graphene bilayer from the electron-hole plasma to optical modes is calculated in the Supplementary Information.
The main change is a suppression in the heat transfer, due to interference effects in the electron-phonon matrix element, partially compensated by an increase in the electronic density of states. 
Note that, for a given power, the absorption is $\alpha \approx 2 \pi / 137$, twice the absorption of a monolayer. 
Results are shown in Fig.~[\ref{fig:bilayer}]. The temperature of the electron-hole plasma, for a given laser power, is significantly increased in bilayer graphene.

{\it Graphene stacks with more than two layers.}
In a system with $N$ layers, the absorbed energy per unit time is $\partial {\cal Q} / \partial t = N \alpha W$, distributed over the $N$ layers, so that the energy absorbed per layer does not change.
In order to estimate the electronic thermal conductivity, we make use of the fact that the low energy band structure of multilayered graphene can be described as a combination of quadratic and linear bands touching at the Dirac point\cite{GNP06}. 
The resulting electronic thermal conductivity is determined by the quadratic bands. 
Its temperature dependence is  $\kappa_{e-h} ( N ) \approx [ k_B^3 T^2 \tau_N ( T ) ]/ ( \hbar^2 a )$, where we assume that, at high temperatures, the scattering time, $\tau_N ( T )$, is determined by electron-electron interactions. 
These interactions couple with similar strength an electron in a given layer and electron-hole pairs in any layer.
Combining this result with the criticality provided by the band touching, we obtain $\tau_N ( T ) \sim \hbar / ( N k_B T )$, so that $\kappa_{e-h} ( N ) \sim ( k_B^2 T ) / ( N \hbar a )$. 
Finally, we obtain that the electron temperature scales as $T ( N ) \propto \sqrt{N}$.

\paragraph{\textbf{Conclusions}---}

We have estimated the electron temperature in a graphene layer under laser irradiation. The temperature is determined by a balance between the power input from the laser, the transfer of energy to optical phonons, and the conduction of heat away from the irradiated region. Temperatures in the order of 1000-2000 K can be reached for a laser power $W \approx 5$ mWatt in regions of radius $R \approx 1 \mu$m. Similar, or higher temperatures can be expected in multilayer stacks. The weak coupling between electrons and acoustic phonons, and the large heat conductivity of these phonons imply that the lattice temperature changes only slightly.

The electron temperature leads to changes in the lattice constant of graphene, even if the lattice temperature does not vary. We find that strains of order $\epsilon \approx 0.01 - 0.02 \%$ are likely. These strains can induce a significant buckling in a suspended sample. Our analysis suggests that light can be used to modify the structural properties of graphene and other two dimensional materials.

\paragraph{\textbf{Acknowledgments}---}
This work was supported by funding from the European Commission under the Graphene Flagship, contract CNECTICT-604391.
Numerical calculations presented in this paper have been performed on a supercomputing system in the Supercomputing Center of Wuhan University.
We thank Thomas Frederiksen for fruitful discussions.

\bibliography{graphene-nourl,dft-nourl}

\begin{thebibliography}{45}
\expandafter\ifx\csname natexlab\endcsname\relax\def\natexlab#1{#1}\fi
\expandafter\ifx\csname bibnamefont\endcsname\relax
  \def\bibnamefont#1{#1}\fi
\expandafter\ifx\csname bibfnamefont\endcsname\relax
  \def\bibfnamefont#1{#1}\fi
\expandafter\ifx\csname citenamefont\endcsname\relax
  \def\citenamefont#1{#1}\fi
\expandafter\ifx\csname url\endcsname\relax
  \def\url#1{\texttt{#1}}\fi
\expandafter\ifx\csname urlprefix\endcsname\relax\def\urlprefix{URL }\fi
\providecommand{\bibinfo}[2]{#2}
\providecommand{\eprint}[2][]{\url{#2}}

\bibitem[{\citenamefont{Amorim et~al.}(2016)\citenamefont{Amorim, Cortijo,
  de~Juan, Grushin, Guinea, Guti\'errez-Rubio, Ochoa, Parente, Rold\'an,
  San-Jose et~al.}}]{Aetal16}
\bibinfo{author}{\bibfnamefont{B.}~\bibnamefont{Amorim}},
  \bibinfo{author}{\bibfnamefont{A.}~\bibnamefont{Cortijo}},
  \bibinfo{author}{\bibfnamefont{F.}~\bibnamefont{de~Juan}},
  \bibinfo{author}{\bibfnamefont{A.~G.} \bibnamefont{Grushin}},
  \bibinfo{author}{\bibfnamefont{F.}~\bibnamefont{Guinea}},
  \bibinfo{author}{\bibfnamefont{A.}~\bibnamefont{Guti\'errez-Rubio}},
  \bibinfo{author}{\bibfnamefont{H.}~\bibnamefont{Ochoa}},
  \bibinfo{author}{\bibfnamefont{V.}~\bibnamefont{Parente}},
  \bibinfo{author}{\bibfnamefont{R.}~\bibnamefont{Rold\'an}},
  \bibinfo{author}{\bibfnamefont{P.}~\bibnamefont{San-Jose}},
  \bibnamefont{et~al.}, \bibinfo{journal}{Physics Reports}
  \textbf{\bibinfo{volume}{617}}, \bibinfo{pages}{1} (\bibinfo{year}{2016}).

\bibitem[{\citenamefont{Rold{\'{a}}n et~al.}(2017)\citenamefont{Rold{\'{a}}n,
  Chirolli, Prada, Silva-Guill{\'{e}}n, San-Jose, and
  Guinea}}]{RoldanTheory2017}
\bibinfo{author}{\bibfnamefont{R.}~\bibnamefont{Rold{\'{a}}n}},
  \bibinfo{author}{\bibfnamefont{L.}~\bibnamefont{Chirolli}},
  \bibinfo{author}{\bibfnamefont{E.}~\bibnamefont{Prada}},
  \bibinfo{author}{\bibfnamefont{J.~A.} \bibnamefont{Silva-Guill{\'{e}}n}},
  \bibinfo{author}{\bibfnamefont{P.}~\bibnamefont{San-Jose}}, \bibnamefont{and}
  \bibinfo{author}{\bibfnamefont{F.}~\bibnamefont{Guinea}},
  \bibinfo{journal}{Chemical Society Reviews} \textbf{\bibinfo{volume}{46}},
  \bibinfo{pages}{4387} (\bibinfo{year}{2017}).

\bibitem[{\citenamefont{Pereira and Castro~Neto}(2009)}]{PN09}
\bibinfo{author}{\bibfnamefont{V.~M.} \bibnamefont{Pereira}} \bibnamefont{and}
  \bibinfo{author}{\bibfnamefont{A.~H.} \bibnamefont{Castro~Neto}},
  \bibinfo{journal}{Phys. Rev. Lett.} \textbf{\bibinfo{volume}{103}},
  \bibinfo{pages}{046801} (\bibinfo{year}{2009}).

\bibitem[{\citenamefont{Mousavi et~al.}(2014)\citenamefont{Mousavi, Rakich, and
  Wang}}]{Metal14c}
\bibinfo{author}{\bibfnamefont{S.~H.} \bibnamefont{Mousavi}},
  \bibinfo{author}{\bibfnamefont{P.~T.} \bibnamefont{Rakich}},
  \bibnamefont{and} \bibinfo{author}{\bibfnamefont{Z.}~\bibnamefont{Wang}},
  \bibinfo{journal}{ACS photonics} \textbf{\bibinfo{volume}{1}},
  \bibinfo{pages}{1107} (\bibinfo{year}{2014}).

\bibitem[{\citenamefont{Salary et~al.}(2016)\citenamefont{Salary, Inampudi,
  Zhang, Tadmor, and Mosallaei}}]{SISZTM16}
\bibinfo{author}{\bibfnamefont{M.~M.} \bibnamefont{Salary}},
  \bibinfo{author}{\bibfnamefont{S.}~\bibnamefont{Inampudi}},
  \bibinfo{author}{\bibfnamefont{K.}~\bibnamefont{Zhang}},
  \bibinfo{author}{\bibfnamefont{E.~B.} \bibnamefont{Tadmor}},
  \bibnamefont{and}
  \bibinfo{author}{\bibfnamefont{H.}~\bibnamefont{Mosallaei}},
  \bibinfo{journal}{Phys. Rev. B} \textbf{\bibinfo{volume}{94}},
  \bibinfo{pages}{235403} (\bibinfo{year}{2016}).

\bibitem[{\citenamefont{Zhang et~al.}(2015)\citenamefont{Zhang, Chang, Wu,
  Xiao, Yi, Lu, Ma, Huang, Zhao, Yan et~al.}}]{Zetal15}
\bibinfo{author}{\bibfnamefont{T.}~\bibnamefont{Zhang}},
  \bibinfo{author}{\bibfnamefont{H.}~\bibnamefont{Chang}},
  \bibinfo{author}{\bibfnamefont{Y.}~\bibnamefont{Wu}},
  \bibinfo{author}{\bibfnamefont{P.}~\bibnamefont{Xiao}},
  \bibinfo{author}{\bibfnamefont{N.}~\bibnamefont{Yi}},
  \bibinfo{author}{\bibfnamefont{Y.}~\bibnamefont{Lu}},
  \bibinfo{author}{\bibfnamefont{Y.}~\bibnamefont{Ma}},
  \bibinfo{author}{\bibfnamefont{Y.}~\bibnamefont{Huang}},
  \bibinfo{author}{\bibfnamefont{K.}~\bibnamefont{Zhao}},
  \bibinfo{author}{\bibfnamefont{X.-Q.} \bibnamefont{Yan}},
  \bibnamefont{et~al.}, \bibinfo{journal}{Nature Photonics}
  \textbf{\bibinfo{volume}{9}}, \bibinfo{pages}{471} (\bibinfo{year}{2015}).

\bibitem[{\citenamefont{Winzer et~al.}(2010)\citenamefont{Winzer, Knorr, and
  Malic}}]{WinzerGraphene2010}
\bibinfo{author}{\bibfnamefont{T.}~\bibnamefont{Winzer}},
  \bibinfo{author}{\bibfnamefont{A.}~\bibnamefont{Knorr}}, \bibnamefont{and}
  \bibinfo{author}{\bibfnamefont{E.}~\bibnamefont{Malic}},
  \bibinfo{journal}{Nano Letters} \textbf{\bibinfo{volume}{10}},
  \bibinfo{pages}{4839} (\bibinfo{year}{2010}), \eprint{1008.1904}.

\bibitem[{\citenamefont{Lui et~al.}(2010)\citenamefont{Lui, Mak, Shan, and
  Heinz}}]{Letal10}
\bibinfo{author}{\bibfnamefont{C.~H.} \bibnamefont{Lui}},
  \bibinfo{author}{\bibfnamefont{K.~F.} \bibnamefont{Mak}},
  \bibinfo{author}{\bibfnamefont{J.}~\bibnamefont{Shan}}, \bibnamefont{and}
  \bibinfo{author}{\bibfnamefont{T.~F.} \bibnamefont{Heinz}},
  \bibinfo{journal}{Phys. Rev. Lett.} \textbf{\bibinfo{volume}{105}},
  \bibinfo{pages}{127404} (\bibinfo{year}{2010}).

\bibitem[{\citenamefont{Johannsen et~al.}(2013)\citenamefont{Johannsen,
  Ulstrup, Cilento, Crepaldi, Zacchigna, Cacho, Turcu, Springate, Fromm, Raidel
  et~al.}}]{Jetal13}
\bibinfo{author}{\bibfnamefont{J.~C.} \bibnamefont{Johannsen}},
  \bibinfo{author}{\bibfnamefont{S.}~\bibnamefont{Ulstrup}},
  \bibinfo{author}{\bibfnamefont{F.}~\bibnamefont{Cilento}},
  \bibinfo{author}{\bibfnamefont{A.}~\bibnamefont{Crepaldi}},
  \bibinfo{author}{\bibfnamefont{M.}~\bibnamefont{Zacchigna}},
  \bibinfo{author}{\bibfnamefont{C.}~\bibnamefont{Cacho}},
  \bibinfo{author}{\bibfnamefont{I.~C.~E.} \bibnamefont{Turcu}},
  \bibinfo{author}{\bibfnamefont{E.}~\bibnamefont{Springate}},
  \bibinfo{author}{\bibfnamefont{F.}~\bibnamefont{Fromm}},
  \bibinfo{author}{\bibfnamefont{C.}~\bibnamefont{Raidel}},
  \bibnamefont{et~al.}, \bibinfo{journal}{Phys. Rev. Lett.}
  \textbf{\bibinfo{volume}{111}}, \bibinfo{pages}{027403}
  (\bibinfo{year}{2013}).

\bibitem[{\citenamefont{Gierz et~al.}(2013)\citenamefont{Gierz, Petersen,
  Mitrano, Cacho, Turcu, Springate, St{\"o}hr, K{\"o}hler, Starke, and
  Cavalleri}}]{Getal13}
\bibinfo{author}{\bibfnamefont{I.}~\bibnamefont{Gierz}},
  \bibinfo{author}{\bibfnamefont{J.~C.} \bibnamefont{Petersen}},
  \bibinfo{author}{\bibfnamefont{M.}~\bibnamefont{Mitrano}},
  \bibinfo{author}{\bibfnamefont{C.}~\bibnamefont{Cacho}},
  \bibinfo{author}{\bibfnamefont{E.}~\bibnamefont{Turcu}},
  \bibinfo{author}{\bibfnamefont{E.}~\bibnamefont{Springate}},
  \bibinfo{author}{\bibfnamefont{A.}~\bibnamefont{St{\"o}hr}},
  \bibinfo{author}{\bibfnamefont{A.}~\bibnamefont{K{\"o}hler}},
  \bibinfo{author}{\bibfnamefont{U.}~\bibnamefont{Starke}}, \bibnamefont{and}
  \bibinfo{author}{\bibfnamefont{A.}~\bibnamefont{Cavalleri}},
  \bibinfo{journal}{Nature Materials} \textbf{\bibinfo{volume}{12}},
  \bibinfo{pages}{1119} (\bibinfo{year}{2013}).

\bibitem[{\citenamefont{Johannsen et~al.}(2015)\citenamefont{Johannsen,
  Ulstrup, Crepaldi, Cilento, Zacchigna, Miwa, Cacho, Chapman, Springate, Fromm
  et~al.}}]{Jetal15}
\bibinfo{author}{\bibfnamefont{J.~C.} \bibnamefont{Johannsen}},
  \bibinfo{author}{\bibfnamefont{S.}~\bibnamefont{Ulstrup}},
  \bibinfo{author}{\bibfnamefont{A.}~\bibnamefont{Crepaldi}},
  \bibinfo{author}{\bibfnamefont{F.}~\bibnamefont{Cilento}},
  \bibinfo{author}{\bibfnamefont{M.}~\bibnamefont{Zacchigna}},
  \bibinfo{author}{\bibfnamefont{J.~A.} \bibnamefont{Miwa}},
  \bibinfo{author}{\bibfnamefont{C.}~\bibnamefont{Cacho}},
  \bibinfo{author}{\bibfnamefont{v.}~\bibnamefont{Chapman}},
  \bibinfo{author}{\bibfnamefont{E.}~\bibnamefont{Springate}},
  \bibinfo{author}{\bibfnamefont{F.}~\bibnamefont{Fromm}},
  \bibnamefont{et~al.}, \bibinfo{journal}{Nano Lett.}
  \textbf{\bibinfo{volume}{15}}, \bibinfo{pages}{326} (\bibinfo{year}{2015}).

\bibitem[{\citenamefont{Ma et~al.}(2016)\citenamefont{Ma, Andersen, Nair,
  Gabor, Massicotte, Lui, Young, Fang, Watanabe, Taniguchi et~al.}}]{Metal16}
\bibinfo{author}{\bibfnamefont{Q.}~\bibnamefont{Ma}},
  \bibinfo{author}{\bibfnamefont{T.~I.} \bibnamefont{Andersen}},
  \bibinfo{author}{\bibfnamefont{N.~L.} \bibnamefont{Nair}},
  \bibinfo{author}{\bibfnamefont{N.~M.} \bibnamefont{Gabor}},
  \bibinfo{author}{\bibfnamefont{M.}~\bibnamefont{Massicotte}},
  \bibinfo{author}{\bibfnamefont{C.~H.} \bibnamefont{Lui}},
  \bibinfo{author}{\bibfnamefont{A.~F.} \bibnamefont{Young}},
  \bibinfo{author}{\bibfnamefont{W.}~\bibnamefont{Fang}},
  \bibinfo{author}{\bibfnamefont{K.}~\bibnamefont{Watanabe}},
  \bibinfo{author}{\bibfnamefont{T.}~\bibnamefont{Taniguchi}},
  \bibnamefont{et~al.}, \bibinfo{journal}{Nature Physics}
  \textbf{\bibinfo{volume}{12}}, \bibinfo{pages}{455} (\bibinfo{year}{2016}).

\bibitem[{\citenamefont{Bistritzer and MacDonald}(2009)}]{BM09}
\bibinfo{author}{\bibfnamefont{R.}~\bibnamefont{Bistritzer}} \bibnamefont{and}
  \bibinfo{author}{\bibfnamefont{A.~H.} \bibnamefont{MacDonald}},
  \bibinfo{journal}{Phys. Rev. Lett.} \textbf{\bibinfo{volume}{102}},
  \bibinfo{pages}{206410} (\bibinfo{year}{2009}).

\bibitem[{\citenamefont{Tse and Das~Sarma}(2009)}]{WS09}
\bibinfo{author}{\bibfnamefont{W.-K.} \bibnamefont{Tse}} \bibnamefont{and}
  \bibinfo{author}{\bibfnamefont{S.}~\bibnamefont{Das~Sarma}},
  \bibinfo{journal}{Phys. Rev. B} \textbf{\bibinfo{volume}{79}},
  \bibinfo{pages}{235406} (\bibinfo{year}{2009}).

\bibitem[{\citenamefont{Song et~al.}(2012)\citenamefont{Song, Reizer, and
  Levitov}}]{SRL12}
\bibinfo{author}{\bibfnamefont{J.~C.~W.} \bibnamefont{Song}},
  \bibinfo{author}{\bibfnamefont{M.~Y.} \bibnamefont{Reizer}},
  \bibnamefont{and} \bibinfo{author}{\bibfnamefont{L.~S.}
  \bibnamefont{Levitov}}, \bibinfo{journal}{Phys. Rev. Lett.}
  \textbf{\bibinfo{volume}{109}}, \bibinfo{pages}{106602}
  (\bibinfo{year}{2012}).

\bibitem[{\citenamefont{Song and Levitov}(2015)}]{SL15}
\bibinfo{author}{\bibfnamefont{J.~C.~W.} \bibnamefont{Song}} \bibnamefont{and}
  \bibinfo{author}{\bibfnamefont{L.~S.} \bibnamefont{Levitov}},
  \bibinfo{journal}{J. Phys.: Condens. Matter} \textbf{\bibinfo{volume}{27}},
  \bibinfo{pages}{164201} (\bibinfo{year}{2015}).

\bibitem[{\citenamefont{Gabor et~al.}(2011)\citenamefont{Gabor, Song, Ma, Nair,
  Taychatanapat, Watanabe, Taniguchi, Levitov, and Jarillo-Herrero}}]{Getal11}
\bibinfo{author}{\bibfnamefont{N.~M.} \bibnamefont{Gabor}},
  \bibinfo{author}{\bibfnamefont{J.~C.~W.} \bibnamefont{Song}},
  \bibinfo{author}{\bibfnamefont{Q.}~\bibnamefont{Ma}},
  \bibinfo{author}{\bibfnamefont{N.~L.} \bibnamefont{Nair}},
  \bibinfo{author}{\bibfnamefont{T.}~\bibnamefont{Taychatanapat}},
  \bibinfo{author}{\bibfnamefont{K.}~\bibnamefont{Watanabe}},
  \bibinfo{author}{\bibfnamefont{T.}~\bibnamefont{Taniguchi}},
  \bibinfo{author}{\bibfnamefont{L.~S.} \bibnamefont{Levitov}},
  \bibnamefont{and}
  \bibinfo{author}{\bibfnamefont{P.}~\bibnamefont{Jarillo-Herrero}},
  \bibinfo{journal}{Science} \textbf{\bibinfo{volume}{334}},
  \bibinfo{pages}{648} (\bibinfo{year}{2011}).

\bibitem[{\citenamefont{Tielrooij et~al.}(2013)\citenamefont{Tielrooij, Song,
  Jensen, Centeno, Pesquera, {Zurutuza Elorza}, Bonn, Levitov, and
  Koppens}}]{Tetal13}
\bibinfo{author}{\bibfnamefont{K.~J.} \bibnamefont{Tielrooij}},
  \bibinfo{author}{\bibfnamefont{J.~C.~W.} \bibnamefont{Song}},
  \bibinfo{author}{\bibfnamefont{S.~A.} \bibnamefont{Jensen}},
  \bibinfo{author}{\bibfnamefont{A.}~\bibnamefont{Centeno}},
  \bibinfo{author}{\bibfnamefont{A.}~\bibnamefont{Pesquera}},
  \bibinfo{author}{\bibfnamefont{A.}~\bibnamefont{{Zurutuza Elorza}}},
  \bibinfo{author}{\bibfnamefont{M.}~\bibnamefont{Bonn}},
  \bibinfo{author}{\bibfnamefont{L.~S.} \bibnamefont{Levitov}},
  \bibnamefont{and} \bibinfo{author}{\bibfnamefont{F.~H.~L.}
  \bibnamefont{Koppens}}, \bibinfo{journal}{Nature Phys.}
  \textbf{\bibinfo{volume}{9}}, \bibinfo{pages}{248} (\bibinfo{year}{2013}).

\bibitem[{\citenamefont{Song et~al.}(2013)\citenamefont{Song, Tielrooij,
  Koppens, and Levitov}}]{STKL13}
\bibinfo{author}{\bibfnamefont{J.~C.~W.} \bibnamefont{Song}},
  \bibinfo{author}{\bibfnamefont{K.~J.} \bibnamefont{Tielrooij}},
  \bibinfo{author}{\bibfnamefont{F.~H.~L.} \bibnamefont{Koppens}},
  \bibnamefont{and} \bibinfo{author}{\bibfnamefont{L.~S.}
  \bibnamefont{Levitov}}, \bibinfo{journal}{Phys. Rev. B}
  \textbf{\bibinfo{volume}{87}}, \bibinfo{pages}{155429}
  (\bibinfo{year}{2013}).

\bibitem[{\citenamefont{Tielrooij et~al.}(2015)\citenamefont{Tielrooij,
  Piatkowski, Massicotte, Woessner, Ma, Lee, Myhro, Lau, Jarillo-Herrero, van
  Hulst et~al.}}]{Tetal15}
\bibinfo{author}{\bibfnamefont{K.~J.} \bibnamefont{Tielrooij}},
  \bibinfo{author}{\bibfnamefont{L.}~\bibnamefont{Piatkowski}},
  \bibinfo{author}{\bibfnamefont{M.}~\bibnamefont{Massicotte}},
  \bibinfo{author}{\bibfnamefont{A.}~\bibnamefont{Woessner}},
  \bibinfo{author}{\bibfnamefont{Q.}~\bibnamefont{Ma}},
  \bibinfo{author}{\bibfnamefont{Y.}~\bibnamefont{Lee}},
  \bibinfo{author}{\bibfnamefont{K.~S.} \bibnamefont{Myhro}},
  \bibinfo{author}{\bibfnamefont{C.~N.} \bibnamefont{Lau}},
  \bibinfo{author}{\bibfnamefont{P.}~\bibnamefont{Jarillo-Herrero}},
  \bibinfo{author}{\bibfnamefont{N.~F.} \bibnamefont{van Hulst}},
  \bibnamefont{et~al.}, \bibinfo{journal}{Nature Nanotech.}
  \textbf{\bibinfo{volume}{10}}, \bibinfo{pages}{437} (\bibinfo{year}{2015}).

\bibitem[{\citenamefont{Mazzola et~al.}(2013)\citenamefont{Mazzola, Wells,
  Yakimova, Ulstrup, Miwa, Balog, Bianchi, Leandersson, Adell, Hofmann
  et~al.}}]{Metal13}
\bibinfo{author}{\bibfnamefont{F.}~\bibnamefont{Mazzola}},
  \bibinfo{author}{\bibfnamefont{J.~W.} \bibnamefont{Wells}},
  \bibinfo{author}{\bibfnamefont{R.}~\bibnamefont{Yakimova}},
  \bibinfo{author}{\bibfnamefont{S.}~\bibnamefont{Ulstrup}},
  \bibinfo{author}{\bibfnamefont{J.~A.} \bibnamefont{Miwa}},
  \bibinfo{author}{\bibfnamefont{R.}~\bibnamefont{Balog}},
  \bibinfo{author}{\bibfnamefont{M.}~\bibnamefont{Bianchi}},
  \bibinfo{author}{\bibfnamefont{M.}~\bibnamefont{Leandersson}},
  \bibinfo{author}{\bibfnamefont{J.}~\bibnamefont{Adell}},
  \bibinfo{author}{\bibfnamefont{P.}~\bibnamefont{Hofmann}},
  \bibnamefont{et~al.}, \bibinfo{journal}{Phys. Rev. Lett.}
  \textbf{\bibinfo{volume}{111}}, \bibinfo{pages}{216806}
  (\bibinfo{year}{2013}).

\bibitem[{\citenamefont{Mazzola et~al.}(2017)\citenamefont{Mazzola,
  Frederiksen, Balasubramanian, Hofmann, Hellsing, and Wells}}]{MFBHHW17}
\bibinfo{author}{\bibfnamefont{F.}~\bibnamefont{Mazzola}},
  \bibinfo{author}{\bibfnamefont{T.}~\bibnamefont{Frederiksen}},
  \bibinfo{author}{\bibfnamefont{T.}~\bibnamefont{Balasubramanian}},
  \bibinfo{author}{\bibfnamefont{P.}~\bibnamefont{Hofmann}},
  \bibinfo{author}{\bibfnamefont{B.}~\bibnamefont{Hellsing}}, \bibnamefont{and}
  \bibinfo{author}{\bibfnamefont{J.~W.} \bibnamefont{Wells}},
  \bibinfo{journal}{Phys. Rev. B} \textbf{\bibinfo{volume}{95}},
  \bibinfo{pages}{075430} (\bibinfo{year}{2017}).

\bibitem[{si()}]{si}
\bibinfo{note}{See Supplementary Information.}

\bibitem[{\citenamefont{Hamm et~al.}(2016)\citenamefont{Hamm, Page, Bravo-Abad,
  Garcia-Vidal, and Hess}}]{HPBGH16}
\bibinfo{author}{\bibfnamefont{J.~M.} \bibnamefont{Hamm}},
  \bibinfo{author}{\bibfnamefont{A.~F.} \bibnamefont{Page}},
  \bibinfo{author}{\bibfnamefont{J.}~\bibnamefont{Bravo-Abad}},
  \bibinfo{author}{\bibfnamefont{F.~J.} \bibnamefont{Garcia-Vidal}},
  \bibnamefont{and} \bibinfo{author}{\bibfnamefont{O.}~\bibnamefont{Hess}},
  \bibinfo{journal}{Phys. Rev. B} \textbf{\bibinfo{volume}{93}},
  \bibinfo{pages}{041408} (\bibinfo{year}{2016}).

\bibitem[{sub()}]{substrate}
\bibinfo{note}{Note that, as we are considering a suspended system, we do not
  include substrate modes\cite{LPRA12}.}

\bibitem[{\citenamefont{Balandin et~al.}(2008)\citenamefont{Balandin, Ghosh,
  Bao, Calizo, Teweldebrhan, Miao, and Lau}}]{Betal08}
\bibinfo{author}{\bibfnamefont{A.~A.} \bibnamefont{Balandin}},
  \bibinfo{author}{\bibfnamefont{S.}~\bibnamefont{Ghosh}},
  \bibinfo{author}{\bibfnamefont{W.}~\bibnamefont{Bao}},
  \bibinfo{author}{\bibfnamefont{I.}~\bibnamefont{Calizo}},
  \bibinfo{author}{\bibfnamefont{D.}~\bibnamefont{Teweldebrhan}},
  \bibinfo{author}{\bibfnamefont{F.}~\bibnamefont{Miao}}, \bibnamefont{and}
  \bibinfo{author}{\bibfnamefont{C.~N.} \bibnamefont{Lau}},
  \bibinfo{journal}{Nano Lett.} \textbf{\bibinfo{volume}{8}},
  \bibinfo{pages}{902} (\bibinfo{year}{2008}).

\bibitem[{\citenamefont{Baladin}(2011)}]{B11}
\bibinfo{author}{\bibfnamefont{A.~A.} \bibnamefont{Baladin}},
  \bibinfo{journal}{Nature Materials} \textbf{\bibinfo{volume}{10}},
  \bibinfo{pages}{569} (\bibinfo{year}{2011}).

\bibitem[{\citenamefont{Yadav et~al.}(2019)\citenamefont{Yadav, Trushin, and
  Pauly}}]{YTP19}
\bibinfo{author}{\bibfnamefont{D.}~\bibnamefont{Yadav}},
  \bibinfo{author}{\bibfnamefont{M.}~\bibnamefont{Trushin}}, \bibnamefont{and}
  \bibinfo{author}{\bibfnamefont{F.}~\bibnamefont{Pauly}},
  \bibinfo{journal}{Phys. Rev. B} \textbf{\bibinfo{volume}{99}},
  \bibinfo{pages}{155410} (\bibinfo{year}{2019}).

\bibitem[{\citenamefont{Nair et~al.}(2008)\citenamefont{Nair, Blake,
  Grigorenko, Novoselov, Booth, Stauber, Peres, and Geim}}]{Netal08}
\bibinfo{author}{\bibfnamefont{R.~R.} \bibnamefont{Nair}},
  \bibinfo{author}{\bibfnamefont{P.}~\bibnamefont{Blake}},
  \bibinfo{author}{\bibfnamefont{A.~N.} \bibnamefont{Grigorenko}},
  \bibinfo{author}{\bibfnamefont{K.~S.} \bibnamefont{Novoselov}},
  \bibinfo{author}{\bibfnamefont{T.~J.} \bibnamefont{Booth}},
  \bibinfo{author}{\bibfnamefont{T.}~\bibnamefont{Stauber}},
  \bibinfo{author}{\bibfnamefont{N.~M.~R.} \bibnamefont{Peres}},
  \bibnamefont{and} \bibinfo{author}{\bibfnamefont{A.~K.} \bibnamefont{Geim}},
  \bibinfo{journal}{Science} \textbf{\bibinfo{volume}{320}},
  \bibinfo{pages}{1308} (\bibinfo{year}{2008}).

\bibitem[{\citenamefont{Hohenberg and Kohn}(1964)}]{HohKoh1964}
\bibinfo{author}{\bibfnamefont{P.}~\bibnamefont{Hohenberg}} \bibnamefont{and}
  \bibinfo{author}{\bibfnamefont{W.}~\bibnamefont{Kohn}},
  \bibinfo{journal}{Physical Review} \textbf{\bibinfo{volume}{136}},
  \bibinfo{pages}{B864} (\bibinfo{year}{1964}).

\bibitem[{\citenamefont{Kohn and Sham}(1965)}]{KohSha1965}
\bibinfo{author}{\bibfnamefont{W.}~\bibnamefont{Kohn}} \bibnamefont{and}
  \bibinfo{author}{\bibfnamefont{L.~J.} \bibnamefont{Sham}},
  \bibinfo{journal}{Physical Review} \textbf{\bibinfo{volume}{140}},
  \bibinfo{pages}{A1133} (\bibinfo{year}{1965}).

\bibitem[{\citenamefont{Soler et~al.}(2002)\citenamefont{Soler, Artacho, Gale,
  Garc\'ia, Junquera, Ordej\'on, and S\'anchez-Portal}}]{SolArt2002}
\bibinfo{author}{\bibfnamefont{J.~M.} \bibnamefont{Soler}},
  \bibinfo{author}{\bibfnamefont{E.}~\bibnamefont{Artacho}},
  \bibinfo{author}{\bibfnamefont{J.~D.} \bibnamefont{Gale}},
  \bibinfo{author}{\bibfnamefont{A.}~\bibnamefont{Garc\'ia}},
  \bibinfo{author}{\bibfnamefont{J.}~\bibnamefont{Junquera}},
  \bibinfo{author}{\bibfnamefont{P.}~\bibnamefont{Ordej\'on}},
  \bibnamefont{and}
  \bibinfo{author}{\bibfnamefont{D.}~\bibnamefont{S\'anchez-Portal}},
  \bibinfo{journal}{Journal of Physics: Condensed Matter}
  \textbf{\bibinfo{volume}{14}}, \bibinfo{pages}{2745} (\bibinfo{year}{2002}).

\bibitem[{\citenamefont{Artacho et~al.}(2008)\citenamefont{Artacho, Anglada,
  Di\'eguez, Gale, Garc\'ia, Junquera, Martin, Ordej\'on, Pruneda,
  S\'anchez-Portal et~al.}}]{ArtAng2008}
\bibinfo{author}{\bibfnamefont{E.}~\bibnamefont{Artacho}},
  \bibinfo{author}{\bibfnamefont{E.}~\bibnamefont{Anglada}},
  \bibinfo{author}{\bibfnamefont{O.}~\bibnamefont{Di\'eguez}},
  \bibinfo{author}{\bibfnamefont{J.~D.} \bibnamefont{Gale}},
  \bibinfo{author}{\bibfnamefont{A.}~\bibnamefont{Garc\'ia}},
  \bibinfo{author}{\bibfnamefont{J.}~\bibnamefont{Junquera}},
  \bibinfo{author}{\bibfnamefont{R.~M.} \bibnamefont{Martin}},
  \bibinfo{author}{\bibfnamefont{P.}~\bibnamefont{Ordej\'on}},
  \bibinfo{author}{\bibfnamefont{J.~M.} \bibnamefont{Pruneda}},
  \bibinfo{author}{\bibfnamefont{D.}~\bibnamefont{S\'anchez-Portal}},
  \bibnamefont{et~al.}, \bibinfo{journal}{Journal of Physics: Condensed Matter}
  \textbf{\bibinfo{volume}{20}}, \bibinfo{pages}{064208}
  (\bibinfo{year}{2008}).

\bibitem[{\citenamefont{Perdew et~al.}(1996)\citenamefont{Perdew, Burke, and
  Ernzerhof}}]{PBE96}
\bibinfo{author}{\bibfnamefont{J.~P.} \bibnamefont{Perdew}},
  \bibinfo{author}{\bibfnamefont{K.}~\bibnamefont{Burke}}, \bibnamefont{and}
  \bibinfo{author}{\bibfnamefont{M.}~\bibnamefont{Ernzerhof}},
  \bibinfo{journal}{Physical Review Letters} \textbf{\bibinfo{volume}{77}},
  \bibinfo{pages}{3865} (\bibinfo{year}{1996}).

\bibitem[{\citenamefont{Troullier and Martins}(1991)}]{tro91}
\bibinfo{author}{\bibfnamefont{N.}~\bibnamefont{Troullier}} \bibnamefont{and}
  \bibinfo{author}{\bibfnamefont{J.~L.} \bibnamefont{Martins}},
  \bibinfo{journal}{Physical Review B} \textbf{\bibinfo{volume}{43}},
  \bibinfo{pages}{1993} (\bibinfo{year}{1991}).

\bibitem[{\citenamefont{Kleinman and Bylander}(1982)}]{klby82}
\bibinfo{author}{\bibfnamefont{L.}~\bibnamefont{Kleinman}} \bibnamefont{and}
  \bibinfo{author}{\bibfnamefont{D.~M.} \bibnamefont{Bylander}},
  \bibinfo{journal}{Physical Review Letters} \textbf{\bibinfo{volume}{48}},
  \bibinfo{pages}{1425} (\bibinfo{year}{1982}).

\bibitem[{\citenamefont{Louie et~al.}(1982)\citenamefont{Louie, Froyen, and
  Cohen}}]{LFC82}
\bibinfo{author}{\bibfnamefont{S.~G.} \bibnamefont{Louie}},
  \bibinfo{author}{\bibfnamefont{S.}~\bibnamefont{Froyen}}, \bibnamefont{and}
  \bibinfo{author}{\bibfnamefont{M.~L.} \bibnamefont{Cohen}},
  \bibinfo{journal}{Physical Review B} \textbf{\bibinfo{volume}{26}},
  \bibinfo{pages}{1738} (\bibinfo{year}{1982}).

\bibitem[{\citenamefont{Artacho et~al.}(1999)\citenamefont{Artacho,
  S\'anchez-Portal, Ordej\'on, Garc\'ia, and Soler}}]{arsan99}
\bibinfo{author}{\bibfnamefont{E.}~\bibnamefont{Artacho}},
  \bibinfo{author}{\bibfnamefont{D.}~\bibnamefont{S\'anchez-Portal}},
  \bibinfo{author}{\bibfnamefont{P.}~\bibnamefont{Ordej\'on}},
  \bibinfo{author}{\bibfnamefont{A.}~\bibnamefont{Garc\'ia}}, \bibnamefont{and}
  \bibinfo{author}{\bibfnamefont{J.~M.} \bibnamefont{Soler}},
  \bibinfo{journal}{Physica Status Solidi (b)} \textbf{\bibinfo{volume}{215}},
  \bibinfo{pages}{809} (\bibinfo{year}{1999}).

\bibitem[{\citenamefont{Monkhorst and Pack}(1976)}]{MonPac76}
\bibinfo{author}{\bibfnamefont{H.~J.} \bibnamefont{Monkhorst}}
  \bibnamefont{and} \bibinfo{author}{\bibfnamefont{J.~D.} \bibnamefont{Pack}},
  \bibinfo{journal}{Physical Review B} \textbf{\bibinfo{volume}{13}},
  \bibinfo{pages}{5188} (\bibinfo{year}{1976}).

\bibitem[{\citenamefont{Wentzcovitch et~al.}(1992)\citenamefont{Wentzcovitch,
  Martins, and Allen}}]{WentzcovitchEnergy1992}
\bibinfo{author}{\bibfnamefont{R.~M.} \bibnamefont{Wentzcovitch}},
  \bibinfo{author}{\bibfnamefont{J.~L.} \bibnamefont{Martins}},
  \bibnamefont{and} \bibinfo{author}{\bibfnamefont{P.~B.} \bibnamefont{Allen}},
  \bibinfo{journal}{Phys. Rev. B} \textbf{\bibinfo{volume}{45}},
  \bibinfo{pages}{11372} (\bibinfo{year}{1992}).

\bibitem[{\citenamefont{de~Gironcoli}(1995)}]{GironcoliLattice1995}
\bibinfo{author}{\bibfnamefont{S.}~\bibnamefont{de~Gironcoli}},
  \bibinfo{journal}{Phys. Rev. B} \textbf{\bibinfo{volume}{51}},
  \bibinfo{pages}{6773} (\bibinfo{year}{1995}).

\bibitem[{\citenamefont{Kresse and Furthmüller}(1996)}]{vasp2}
\bibinfo{author}{\bibfnamefont{G.}~\bibnamefont{Kresse}} \bibnamefont{and}
  \bibinfo{author}{\bibfnamefont{J.}~\bibnamefont{Furthmüller}},
  \bibinfo{journal}{Computational Materials Science}
  \textbf{\bibinfo{volume}{6}}, \bibinfo{pages}{15 } (\bibinfo{year}{1996}).

\bibitem[{\citenamefont{Khestanova et~al.}(2016)\citenamefont{Khestanova,
  Guinea, Fumagalli, Geim, and Grigorieva}}]{KGLGG16}
\bibinfo{author}{\bibfnamefont{E.}~\bibnamefont{Khestanova}},
  \bibinfo{author}{\bibfnamefont{F.}~\bibnamefont{Guinea}},
  \bibinfo{author}{\bibfnamefont{L.}~\bibnamefont{Fumagalli}},
  \bibinfo{author}{\bibfnamefont{A.~K.} \bibnamefont{Geim}}, \bibnamefont{and}
  \bibinfo{author}{\bibfnamefont{I.~V.} \bibnamefont{Grigorieva}},
  \bibinfo{journal}{Nature Comm.} \textbf{\bibinfo{volume}{7}},
  \bibinfo{pages}{12587} (\bibinfo{year}{2016}).

\bibitem[{\citenamefont{Guinea et~al.}(2006)\citenamefont{Guinea, Castro~Neto,
  and Peres}}]{GNP06}
\bibinfo{author}{\bibfnamefont{F.}~\bibnamefont{Guinea}},
  \bibinfo{author}{\bibfnamefont{A.~H.} \bibnamefont{Castro~Neto}},
  \bibnamefont{and} \bibinfo{author}{\bibfnamefont{N.~M.~R.}
  \bibnamefont{Peres}}, \bibinfo{journal}{Phys. Rev. B}
  \textbf{\bibinfo{volume}{73}}, \bibinfo{pages}{245426}
  (\bibinfo{year}{2006}).

\bibitem[{\citenamefont{Low et~al.}(2012)\citenamefont{Low, Perebeinos, Kim,
  Freitag, and Avouris}}]{LPRA12}
\bibinfo{author}{\bibfnamefont{T.}~\bibnamefont{Low}},
  \bibinfo{author}{\bibfnamefont{V.}~\bibnamefont{Perebeinos}},
  \bibinfo{author}{\bibfnamefont{R.}~\bibnamefont{Kim}},
  \bibinfo{author}{\bibfnamefont{M.}~\bibnamefont{Freitag}}, \bibnamefont{and}
  \bibinfo{author}{\bibfnamefont{P.}~\bibnamefont{Avouris}},
  \bibinfo{journal}{Phys. Rev. B} \textbf{\bibinfo{volume}{86}},
  \bibinfo{pages}{045413} (\bibinfo{year}{2012}).

\end{thebibliography}


\begin{thebibliography}{1}
\expandafter\ifx\csname natexlab\endcsname\relax\def\natexlab#1{#1}\fi
\expandafter\ifx\csname bibnamefont\endcsname\relax
  \def\bibnamefont#1{#1}\fi
\expandafter\ifx\csname bibfnamefont\endcsname\relax
  \def\bibfnamefont#1{#1}\fi
\expandafter\ifx\csname citenamefont\endcsname\relax
  \def\citenamefont#1{#1}\fi
\expandafter\ifx\csname url\endcsname\relax
  \def\url#1{\texttt{#1}}\fi
\expandafter\ifx\csname urlprefix\endcsname\relax\def\urlprefix{URL }\fi
\providecommand{\bibinfo}[2]{#2}
\providecommand{\eprint}[2][]{\url{#2}}

\bibitem[{\citenamefont{Castro~Neto and Guinea}(2007)}]{NG07}
\bibinfo{author}{\bibfnamefont{A.~H.} \bibnamefont{Castro~Neto}}
  \bibnamefont{and} \bibinfo{author}{\bibfnamefont{F.}~\bibnamefont{Guinea}},
  \bibinfo{journal}{Phys. Rev. B} \textbf{\bibinfo{volume}{75}},
  \bibinfo{pages}{045404} (\bibinfo{year}{2007}).

\end{thebibliography}

\end{document}


\title{Supplementary Information for: ``Electron heating and mechanical properties of graphene''}

\author{J. A. Silva-Guill\'{e}n}
\email{josilgui@gmail.com} 
\affiliation{Key Laboratory of Artificial Micro- and Nano-structures of Ministry of Education and School of Physics and Technology, Wuhan University, Wuhan 430072, China}

\author{F. Guinea}
\affiliation{Fundaci\'on IMDEA Nanociencia, C/Faraday 9, Campus Cantoblanco, 28049 Madrid, Spain}
\affiliation{Department of Physics and Astronomy, University of Manchester, Oxford Road, Manchester M13 9PL, UK}
\affiliation{Donostia International Physics Center (DIPC) -- UPV/EHU, E-20018 San Sebasti\'an, Spain}

\maketitle

%

\section{Radiation pressure on a graphene layer}\label{sec:rad_press}
As a first step to understand the effect of temperature of electrons to the graphene lattice, we can consider the graphene layer as a membrane.
This membrane is suspended over a circular hole in a substrate to which we apply radiation using a laser (see Fig. 1 in the main text).
We will study the effect of the pressure exerted by the radiation to the graphene layer on its structural properties. 
It is important to note that, although we assume a circular shape of the hole, changing its shape will not affect the order of magnitude of the following results.
From the theory of membranes, if we define the radius of the hole as $R$, the deflection at the center as $h_{max}$, and the applied pressure as $P$ we find that:

\begin{equation}
  h_{max} = R\left( \frac{c_VPR}{4c_1Y} \right)^{\frac{1}{3}},
  \label{eq:hmax}
\end{equation}

where $Y$ is the Young modulus of the membrane, and $c_1\approx2.8$ and $c_V\approx1.7$ are numerical constants which depend on the shape of the suspended region. 
We assume that the pressure is only due to the momentum transfer by photons coming from the laser beam which are absorbed by the graphene layer. 
Since both the pressure, $P$, and $Y$ are proportional to the number of layers, $N$, in the suspended region, $h_{max}$ is independent of $N$, and we can study the case of the graphene layer ($N=1$) without loss of generality.

We consider a laser of power $W$, whose emitted electromagnetic radiation is localized within a circular area of radius $R_{spot}$. 
The pressure of the beam can be written as
\begin{equation}
  P = \alpha \frac{d}{d t} \frac{P_{rad}}{A} = \frac{\alpha W}{\pi R_{spot}^2 c},
  \label{eq:press}
\end{equation}
where $P_{rad}$ is the rate of total momentum.

The absorption, $\alpha$, of a graphene monolayer is
\begin{equation}
  \alpha \approx \pi \frac{e^2}{\hbar c} \approx \frac{\pi}{137} \approx 0.023,
\end{equation}
where $e^2 / ( \hbar c ) \approx 1 / 137$ is the fine structure constant.

The pressure, $P$, is proportional to the laser power, $W$, and inversely proportional to the area of the spot where the radiation is concentrated, $A_{spot}=\pi R^{2}_{spot}$.

If we assume $W=1\textrm{mW}$ and $R_{spot}=10\mu\textrm{m}$, which are the typical values used in the experimental setup described above, we find that $P\approx 7.6\cdot10^{-4}$ Pa.
The Young modulus of graphene is $Y\approx 22\textrm{eV} \cdot$ \AA$^{-2}$.
For a suspended are of radius $R=6\mu\textrm{m}$, we obtain $h_{max}\approx1.4\textrm{nm}$.
If we increase the power of the laser to 10 mW, we find that $h_{max}\approx 3 \textrm{nm}$.

\section{Heat transfer to optical phonons.}
We analyze the scattering between electrons and optical phonons using the simplified model used in\cite{NG07}. We use a constant frequency for the phonons, $\omega_{op}$. The coupling between electrons and phonons is due to the modification of the nearest neighbor hopping due to the change in bond length.
\begin{align}
{\cal H}_{e-ph} &= \frac{\partial t}{\partial l} c^\dag_i c_j \left( | \vec{r}_i - \vec{r}_j | -a \right) = \nonumber \\ &= \frac{\beta t}{a} \frac{3}{2} \times \frac{\hbar f_{x,y}}{2 M} \sum_{\vec{k}_1 , \vec{k}_2} c^\dag_{\vec{k}_1}  c_{\vec{k}_2} u^{op , \{ x , y \}}_{\vec{k}_1 - \vec{k}_2}
\label{coupling}
\end{align}
where $t \approx 3$ eV is the nearest neighbor hopping, and $a \approx 1.4$~\AA~is the bond length, and $u_{\vec{q}}^{op, \{ x, y \}}$ is the displacement of an optical mode of wavevector $\vec{q}$ polarized along the $x,y$ axis, and $f_{x,y} = \left\{ \cos ( \phi_{\vec{k}_1 - \vec{k}_2} ) , \sin ( \phi_{\vec{k}_1 - \vec{k}_2} ) \right\}$. We write $\partial t / \partial l = \beta t / a$, where $\beta \approx 2- 3 $ is a dimensionless parameter.

There are two optical modes at the $\Gamma$ point. The power dissipated can be calculated using Fermi's Golden Rule and Eq. (\ref{coupling})
\begin{widetext}
\begin{align}
\frac{\partial Q}{\partial t} &= \frac{4}{( 4 \pi^2 )^2} \frac{9}{4} \frac{\hbar \beta^2 t^2 \Omega}{2 a^2 M} \int k_1 d k_1 d \phi_1 \int k_2 dk_2 d \phi_2 n_{\vec{k}_1} \left( 1 - n_{\vec{k}_2} \right) \delta \left( \hbar v_F | \vec{k}_1 | - \hbar v_F | \vec{k}_2 | - \hbar \omega_{op} \right) = \nonumber \\ &= \frac{27 \sqrt{3}}{8 \pi^2} \frac{\hbar \beta^2 t^2 ( k_B T )^3}{M ( \hbar v_F )^4} \int_{- \infty}^{\infty} dx \frac{ x | x - \tilde{\omega}_{op} | e^{-x}}{( 1 + e^{-x} ) ( 1+ e^{-x + \tilde{\omega}_{op}} )} = \alpha_{op} \frac{( k_B T )^3}{( \hbar v_F )^4} {\cal F} \left( \frac{\hbar \omega_{op}}{k_B T} \right)
\end{align}
\end{widetext}
where $\Omega = 3 \sqrt{3} a^2 / 2$ is the size of the unit cell, $M$ is the mass of the carbon atom, and $n_{\vec{k} }$ is the Fermi-Thomas distribution. The numerator $4$ in the first coefficient in the first expression accounts for the spin and valley degeneracies, and $\tilde{\omega}_{op} = ( \hbar \omega_{op} ) / ( k_B T )$. The function ${\cal F} ( x )$ satisfies $\lim_{x \rightarrow \infty} {\cal F} ( x ) = \pi^2 / 3$.

In a graphene bilayer there is a single parabolic band at energies below the interlayer hopping element, $t_\perp \approx 0.4$ eV. The dispersion of this band is $\epsilon_{\vec k} = ( \hbar v_F | \vec{k} | )^2 / t_\perp$. The electron-phonon coupling becomes
\begin{widetext}
\begin{align}
{\cal H}_{e-ph}^{BLG} &= \frac{\partial t}{\partial l} c^\dag_i c_j \left( | \vec{r}_i - \vec{r}_j | -a \right) = \nonumber \\ &= \frac{\beta t}{a} \frac{v_F}{2 t_\perp} \left( \left| \vec{k}_1 \right| e^{i \phi_1} + \left| \vec{k}_2 \right| e^{i \phi_2} \right) \frac{3}{2} \times \frac{\hbar f_{x,y}}{2 M} \sum_{\vec{k}_1 , \vec{k}_2} c^\dag_{\vec{k}_1}  c_{\vec{k}_2} u^{op , \{ x , y \}}_{\vec{k}_1 - \vec{k}_2}
\label{coupling2}
\end{align}
\end{widetext}
The change in the matrix element with respect to Eq. (\ref{coupling}) is is due to the suppression of intralayer electron hopping at low energies and momenta. Following the same steps as before, and taking into account the doubling in the number of phonon modes, we find
\begin{widetext}
\begin{align}
\frac{\partial Q}{\partial t} &=  \frac{27 \sqrt{3}}{64 \pi^2} \frac{\hbar \beta^2 t^2 t_\perp( k_B T )^2}{M ( \hbar v_F )^4} \int_{- \infty}^{\infty} dx \frac{ ( x + | x- \tilde{\omega}_{op} | ) e^{-x}}{( 1 + e^{-x} ) ( 1+ e^{-x + \tilde{\omega}_{op}} )} = \alpha_{op}^{BLG} \frac{( k_B T )^3}{( \hbar v_F )^4} {\cal F}^{BLG} \left( \frac{\hbar \omega_{op}}{k_B T} \right)
\end{align}
\end{widetext}
where $\lim_{x \rightarrow \infty} {\cal F}^{BLG} ( x ) = 4 \log ( 2 )$.

\bibliography{graphene-nourl}